\documentclass[preprint,preprintnumbers,showpacs,nofootinbib,amsmath,amssymb,aps,prd,letterpaper,
twocolumn
superscriptaddress,
groupedaddress,
unsortedaddress,
runinaddress,
frontmatterverbose,
nobibnotes,
bibnotes,
]{revtex4-1}
\usepackage{graphicx}
\usepackage{dcolumn}
\usepackage{bm}
\usepackage{subfigure}
\usepackage{color}
\usepackage[active]{srcltx}
\usepackage{amsmath,amsfonts,amssymb,amsthm,amstext,amscd,eucal,srcltx}
\usepackage{epsfig,graphicx,bm}
\usepackage{epstopdf, epsf}
\usepackage{dcolumn}
\usepackage{hyperref}
\usepackage{lipsum}
\usepackage{feynmf}
\everymath{\displaystyle}
\begin{document}
\title{Polarization Test of Higgs Spin and Parity}
\author{Firooz Arash}
\thanks{farash@aut.ac.ir}
\affiliation{Physics Department,Tafresh University, Tafresh, Iran.}

\begin{abstract}
A polarization test is applied to determine the spin and the parity of the observed resonance at LHC, which is believed to be the expected "Higgs" particle. The test is based on very general principles and is completely independent of dynamical assumptions. We have also identified a set of observables that discriminate resonances with $J^P=0^+,0^-, 2^-$ and $2^+$. Furthermore, the same set can be used to gain useful and important information on the magnitude of each helicity amplitude contributing to the $gg\rightarrow \gamma\gamma$ process .
\date{\today}
\end{abstract}
\maketitle
\section{INTRODUCTION}
Recently identified resonance with a mass of around $125$ GeV at Large Hadron Collider by ATLAS \cite{1} and CMS \cite{2} collaborations is believed to be the long-sought Higgs particle expected from the Standard Model. The standard model Higgs is a scalar boson with $J^P=0^+$. Since this is the only elementary particle with spin zero, it is crucial to check its internal quantum numbers, its spin and parity, experimentally. Many studies have been performed as to how its spin and parity can be measured at LHC \cite{3}\cite{4}\cite{5}\cite{6}\cite{7}\cite{8}\cite{9}\cite{10}\cite{11}\cite{12}. Most of these studies are concentrated on the decay of a resonance, $X$, to vector boson pairs,$X\rightarrow ZZ, W^+ W^-$, and the decay process $X\rightarrow t\overline{t}$. These approaches mainly look for the dependence of the cross section on the angle $\theta$ between momentum of one of the initial partons and the momentum of one of the decaying particles. It is expected that the cross section behaves differently with respect to $\theta$ depending on the spin of $X$. Given that vector bosons decay into four leptons, the angular distributions of the final state leptons provide information about the spin and the parity of $X$. Another approach is the investigation of the invariant mass distributions of a Higgs with an associated vector boson in the final state. A detailed analysis of the decay amplitudes for various spin state of the resonance is provided in \cite{13}. These approaches are model dependent and one usually makes certain dynamical assumptions on the couplings of Higgs to other particles. Ideally, however, one would prefer to analyze the Higgs' spin in a way that is model independent. In this paper, we would like to provide a model independent analysis of the Higgs spin and parity, using the polarization observables of the final state particles. We will restrict our analysis to $\gamma\gamma$ final state, but the same analysis can also be used for other final states. Our test of course, requires the measurement of photon polarization, a task that is seriously under consideration. However, it is a difficult task because, it requires photon conversion in the detector and the statistics currently are too low. The situation can improve as the available statistics increases. \\
The paper is based on some earlier research on the structure of the scattering matrix to which the author was an active participant. The organization of the paper is as follows. In section II we give a general description of observable-amplitude structure in a two body reaction. Section III deals with amplitude-observable structure under Lorentz invariance alone. Section IV considers the imposition of discrete symmetries in addition to the Lorentz invariance. In section V we offer a set of general, dynamics independent criteria for the formation of a direct channel resonance and discuss its implications on the spin and the parity of that resonance state. Finally, in section VI we summarize our results.

\section{Observables and Amplitudes}
At High energies (even above a few GeV) only the amplitude description of a reaction is practically feasible. Thus, one needs to choose a particular amplitude formalism. For a given reaction, and for a given set of symmetries holding for that reaction, the number of independent amplitudes describing that reaction is the same in any formalism. The number of amplitudes depends only on the values of the spins of the participating particles and the symmetries apply. For example, in the case of $gg\rightarrow \gamma \gamma$ since both the initial and the final states consist of massless spin-one particles, each particle has only two spin states. If only the Lorentz invariance is assumed, there will be 16 independent amplitudes. When both Lorentz invariance and the parity conservation are imposed, the number reduces to 8. Imposition of time reversal and the identical particle constraints on the initial and final states brings down the number of independent amplitudes to 5.  Another example is the Higgs production in association with a vector boson in reactions such as $pp\rightarrow XZ$ and $p\overline{p}\rightarrow XZ$ for which one would have 12 independent complex amplitudes. In these processes the invariant mass distribution of $X+Z$ is used \cite{14} to discriminate $J^P=0^+, 0^-$ and $J^P=2^+$ cases.\\
The spin observables depend linearly on bilinear products of the complex amplitudes, the relationship is given by a large matrix. To achieve economy and simplicity, this matrix should be as close to a diagonal one as possible. Hermicity requirement prohibits complete diagonalization of this matrix. The class of formalisms in which the matrix is as close to diagonal as possible is called "optimal" \cite{15}. In it, this matrix consists of a string of small matrices along the main diagonal and zeroes everywhere else. The size of the small matrices, for any four particle reaction, are 1-by-1, 2-by-2, 4-by-4 or 8-by-8. If one of the four particles has spin zero, there are no 8-by-8 matrices. \\
Optimal formalisms differ from each other in the orientation of the quantization axes for the participating particles. When only Lorentz invariance applies, each of these axes can point in any direction, independent of the orientation of the other three. When parity conservation is also imposed, the orientation for a given of these four axes must be either in the reaction plane (and hence "planar formalism") or perpendicular to the reaction plane (and hence "transversity formalism"). Unless time reversal invariance or identical particle constraints are additionally imposed, however, the four orientation axes can be chosen independent of each other. For photon or for any massless particle, the helicity formalism is the most natural one since the massless particle can have only two polarization states, which are the positive and negative helicity states. \\
In this paper we will consider only formalisms that are "optimal" in the sense explained above. The helicity formalism, which we will utilize, is a special case of the general optimal formalism. \\
The observables referred to above are "primary" type, that is, in them the polarization states of each of the particles in the reaction are specified. In actual practice, one often deals with experiments in which the polarization state of one or several particles is averaged over, or in other words, unpolarized particles are used. Such observables are obtained from the primary ones and are called "secondary" observables.

\section{The observable-amplitude structure with Lorenz invariance only}
The generation of the amplitude and observable relations in the optimal formalism is a standard procedure. Its general formulation is developed a long time ago\cite{16} and illustrated for numerous specific examples\cite{17}. It is, therefore, suffices here to review the notation and then supply the relations themselves. \\
The amplitudes are denoted by D(c, d; a, b) where c, and d denote the spin projections of the final state particles along the quantization axes and a, and b represent that of the initial state particles. \\
With only Lorentz or rotational invariance, these amplitudes form an independent set and the spin orientation direction can be chosen arbitrarily. In the matrices connecting the bilinear products of amplitudes and the experimental observables, for the case of Lorenz invariance alone, all non-zero numbers are $+1$ or $-1$. Imposition of discrete symmetries limits the choice of orientation axes. \\
In this paper we are primarily interested in $gg\rightarrow\gamma \gamma$ reaction, in which all particles involved are massless and spin one. The most suitable frame for describing that reaction is the helicity frame, where the orientation axes are along the momenta of the the particles. In the helicity frame the spin projections can only assume values +1 and -1 and will be denoted simply by $+$ and $-$ in the argument of the amplitudes. \\
In a general reaction $A+B\rightarrow C+D$, the observables are denoted by $\cal{L}$$(uvH_p ,UVH_P ;\xi\omega H_q ,\Xi\Omega H_Q)$, where $u$ and $v$ are the spin indices for particle A, the indices $U$ and $V$ refer to the spin of particle B, the indices $\xi,\omega$ to particle C and finally, the indices $\Xi$ and $\Omega$ refer to particle D. Each of $H_j$ can be either real (R) or imaginary (I), provided that the subscripts of H is +1 or -1, respectively. For a detailed description of the notation see Ref [15]. For the process $gg\rightarrow\gamma \gamma$, in the primary observables, for each of the arguments in $\cal{L}$, we can have four possibilities of $(++)$, $(--)$, $Re(+-)$ and $Im(+-)$. In Table 1 these polarization states of gluon or photon are simply denoted by $\cal{R}$ (for right circular polarization, $++$ state), $L$ (for left circular polarization, $--$ state) and $R$, $I$ (for $Re(+-)$ and $Im(+-)$ states, respectively). States denoted by $Re(+-)$ and $Im(+-)$ correspond to linearly polarized states. In particular, for a photon or a gluon plane polarized in the direction of $\phi=0^{\circ},180^{\circ}$ we get $-Re(+-)$ and polarization in the direction of$\phi=90^{\circ}, 270^{\circ}$, etc. we obtain $+Re(+-)$. Plane polarization in the direction of $\phi=45^{\circ}, 225^{\circ}, etc.$ goes with $+Im(+-)$.  For polarization in the direction of $\phi=135^{\circ},315^{\circ}, etc$ we have $-Im(+-)$. \\
The secondary observables are obtained from the primary ones. They still contain $R$ and $I$, as before, but instead of $++$ and $--$ the unpolarized state is $A=(++)+(--)$ and the circular asymmetry of photon is defined by $\Delta=(++)-(--)$ .

\section{Observable-Amplitude structure under Lorentz Invariance and Discrete Symmetries}
As discussed earlier, under parity conservation the 16 independent amplitudes reduce to 8. Since we are considering helicity, we know that the reduction of amplitudes from 16 to 8 will not occur by the vanishing of the 8 amplitudes, but will occur by pairwise equalities given by
\begin{equation}
D(c,d;a,b)= (-1)^{a+b+c+d}(-1)^{S_A-S_B+S_C-S_D}D(-c,-d ;-a,-b)
\end{equation}
So, under parity transformation, all helicities reverse sign. This in turn will impose restrictions on the observables, which are expressed as the linear combinations of the bilinear products of the amplitudes.\\
Imposition of time reversal invariance requires that the helicity amplitudes to satisfy
\begin{equation}
D(c,d;a,b)= (-1)^{a+b+c+d}(-1)^{S_A+S_B+S_C+S_D}D(a,b ;c,d)
\end{equation}
This relationship reduces the number of helicity amplitudes from 8 to 6. Finally, when particles A=B and C=D, identical particle restriction also applies and provide additional relation among the helicity amplitudes according to
\begin{equation}
D(c,d;a,b)= (-1)^{a+b+c+d}(-1)^{2S_A+2S_C}D(d,c;b,a)
\end{equation}
Thus, reducing the number of independent amplitudes to 5.\\
These restrictions on the amplitudes when carried over to the observables, result in certain relationships among the observables. In the case of identical particle constraint, one gets
\begin{equation}
{\cal{L}}(uvH_p ,UVH_P ;\xi\omega H_q ,\Xi\Omega H_Q)=(-1)^{\xi+u+\Xi+U+\omega+v+\Omega+V}{\cal{L}}(UVH_P ,uvH_p ;\Xi\Omega H_Q ,\xi\omega H_q)
\end{equation}
 If in addition to identical particle constraint, time reversal invariance constraint is also imposed, we will have the following relations among the observables \cite{18}:
 \begin{eqnarray}
\lefteqn{{\cal{L}}(uvH_p ,UVH_P ;\xi\omega H_q ,\Xi\Omega H_Q)= } \nonumber \\
 & & (-1)^{\xi+u+\Xi+U+\omega+v+\Omega+V+\frac{1}{2}(p-q+P-Q)}{\cal{L}}(\xi\omega H_q ,\Xi\Omega H_Q ;uvH_p ,UVH_P).
 \end{eqnarray}
 In what follows, we will utilize these relations among the amplitudes and the observables to discuss how to distinguish between Higgs particle with spin zero and spin 2.
 \section{ Polarization Test of Higgs Resonance In $gg\rightarrow\gamma\gamma$ Reaction and Its Spin and The Parity }
 The reaction under consideration, $gg\rightarrow\gamma\gamma$, conserves parity and exhibits identical particle feature in both the initial and the final states. We further assume time reversal invariance. Under these conditions, and owing to the fact that all particles involved in that reaction are massless, the spin structure of this reaction becomes identical to four spin $\frac{1}{2}$ particle interaction, for which, the amplitude-observable structure is worked out a long time ago \cite{15}\cite{19}\cite{20}. Here I will mention only those observables that are relevant to $gg\rightarrow\gamma\gamma$, with both initial state gluons being unpolarized. \\
 $J$\emph{-Constraints}: The process $gg\rightarrow X_J\rightarrow \gamma\gamma$ is illustrated in Figure 1.
 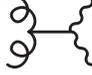
\begin{figure}[h]
\begin{fmffile}{fe}
\begin{fmfgraph*}(40,25)
  \fmfleft{em,ep} \fmflabel{$a$}{ep} \fmflabel{$b$}{em}
   \fmfright{fb,f} \fmflabel{$c$}{fb} \fmflabel{$d$}{f}
  \fmf{gluon}{em,gl,ep}
  \fmf{vanilla}{gl,Z}
  \fmf{photon}{fb,Z,f}
  \fmflabel{$\Gamma_{ab}$}{gl}
  \fmflabel{$\Gamma_{cd}$}{Z}
\end{fmfgraph*}
\end{fmffile}
\caption{\footnotesize   Transverse polarized $He^3$
structure function, $g_2(x,Q^2)$   at $Q^2=5GeV^2$.}
\end{figure}
For a direct channel resonance, if the process proceeds through a single resonance of spin $J$ and all spin quantization axes are in the reaction plane (this choice of direction assures that $L_z =0$), then all amplitudes with $|a-b|>J$ or $|c-d|>J$ must vanish \cite{21}\cite{22}. As mentioned before, a, b, c, and d denote the spin components along some axes for the corresponding reaction $A+B\rightarrow C+D$. We call this as $J$\emph{-Constraints}. \\
If two amplitudes $D(c_1 ,d_1 ;a_1 ,b_1)$ and $D(c_2 ,d_2 ;a_2 ,b_2)$ are both $J$-forbidden, then all eight observables formed from bilinear products of them and permuted amplitudes will vanish. The same is true if the two amplitudes are $J$-allowed but if $a_1+b_1+a_2+b_2=2J+\beta$, where $\beta$ is a positive number. \\
\emph{ Factorization constraint:} A second constraint arises from the $J$-constraint on the two vertices. As discussed in \cite{21}. The interaction depicted in Figure 1 can be visualized as the product of two non-overlapping independent parts, each containing a set of physical particles plus the resonance. Each part is a three particle vertex and at each vertex the number of vertex amplitudes is restricted by the $J$-constraint. A simple three-particle vertex can at most have $N_{Js_1s_2}=(2s_1 +1)(2s_2 +1)(2J+1)$ amplitudes. Then, the number of three-point amplitudes in the overall factorizable four-particle process with an arbitrary $J$- resonance is at most \\
\begin{equation}
N^f =[(2s_1 +1)(2s_2 +1)(2s_3 +1)(2s_4 +1)](2J+1).
\end{equation}
Since the overall reaction has only $N=(2s_1 +1)(2s_2 +1)(2s_3 +1)(2s_4 +1)$ amplitudes, one must have $N^f<N$ . Depending on the value of $J$, this inequality may or may not reduce the number of vertex amplitudes. In fact, for $J>0$ the inequality does not hold and no reduction occurs.\\
The number of three-particle vertex amplitudes enumerated above does not take into account the $J$ constraint. In a real process with $J\rightarrow s_1 +s_2$, then $s_{1z} +s_{2z}$ must not exceed $J$. As shown in \cite{21}\cite{22}, Imposition of this consideration reduces the number of vertex amplitude to
\begin{eqnarray}
N_{J,s_1,s_2}=(2s_1 +1)(2s_2 +1) &  & \mbox{$J\geq s_1+s_2$}  \nonumber  \\
N_{J,s_1,s_2}=(2J+1)(s_1 +s_2 -J)(J+s_2 -s_1 +1)(J-s_2 +s_1 +1) & &\mbox{$s_1+s_2\geq J\geq s_2-s_1$} \hspace {0.05 in} \\
N_{J,s_1,s_2}=(2s_1 +1)(2J +1)  & &\mbox{$s_2-s_1\geq J$}  \nonumber
\end{eqnarray}
This implies that the number of independent amplitudes $N_{J}^{f}$ from factorization constraint and from the $J$-constraints, combined, is reduced to $N_{J}^{f}=N_{J S_1 S_2}+N_{J S_3 S_4}$. \\
For the direct channel resonance, these constraints are most simply expressed for center of mass helicity amplitudes
\begin{equation}
D(c,d;a,b)=\sum_{J} D_{J}(c,d;a,b)d^{J}_{c-d, a-b}(\theta)
\end{equation}
where, $d^{J}_{c-d, a-b}(\theta)$ is the matrix element for rotation by the scattering angle $\theta$ about the normal to the scattering plane. For a resonance of spin $J$ the spin dependence factorizes as follows
\begin{equation}
D(c,d;a,b)\propto\Gamma^{' J}_{cd}\Gamma^{J}_{ab}
\end{equation}
where $\Gamma s$ are proportional to vertex functions having the requisite number of $N_{J S_1 S_2}$ independent components of equation (7). We thus get nonlinear relations among the helicity amplitudes that lead to complicated relations among the observables. However, when parity is conserved, equation (9) simplifies. Then, for spins $s_a$ and $s_b$ and intrinsic parities $\eta_a$ and $\eta_b$ the vertices satisfy
\begin{equation}
\Gamma^{J}_{-a, -b}=\eta_{J}\eta_{a}\eta_{b}(-1)^{s_a+s_b}\Gamma^{J}_{a, b}
\end{equation}
where, $\eta_{J}$ is the naturality of the resonance with spin $J$, i.e. $\eta_{J}=\pm$ for resonance parity $\pm (-1)^J$. So, we get
\begin{equation}
 D(c,d;a,b)=\pm D_J(-c,-d;a,b)
\end{equation}
When the rotation functions are the same on both sides of equation (8), this reduces to a simple relation for the helicity amplitudes. \\
For the reaction $gg\rightarrow \gamma\gamma$ with all pertinent symmetries, as mentioned before, there are only five helicity amplitudes listed below.
\begin{eqnarray}
A_1\equiv D(+,+;+,+) & A_2\equiv D(+,+;+,-)  & A_3\equiv D(+,-;+,-)  \nonumber  \\
A_4\equiv D(+,+;-,-) & A_5 \equiv D(+,-;-+) &
\end{eqnarray}
Applying equations (8) and (10) to these amplitudes leads to the following relations between the pairwise amplitudes for a resonance with $\eta_{J}=\pm 1$,
\begin{eqnarray}
A^J_1=\pm A^J_4, \hspace{1 in}  A^J_3=\pm A^J_5
\end{eqnarray}
These are the signatures for the resonance formation. Since both $A_1$ and $A_4$ go with $d_{00}^{J}(\theta)$, we get
\begin{equation}
A_1=\pm A_4
\end{equation}
With no general restrictions on $A_2$, $A_3$, and $A_5$. For a spin zero resonance, $X_0$, due to $J$-constraint, we also have
\begin{equation}
A_2=A_3=A_5=0.
\end{equation}
So, for a $X_0$ resonance there is only one independent helicity amplitude and as such, all obervables with one or both particle's polarization specified vanish uniquely. Thus, we are left only with the unpolarized cross-section proportional to $|A_1|^2=|A_4|^2$.
Since the amplitude $A_3$ goes with $d^{J}_{11}(\theta)$ while $A_5$ goes with $d^{J}_{1-1}(\theta)$, there is no angle independent relation between them, but their relative phase is zero or $\pi$ for $^3(L=J)$ states. \\
Our amplitude test in general does not prohibit the formation of a resonance state with spin 1. However, the decay of such state into two photons are forbidden by Landau-Yang theorem \cite{23}\cite{24}. It is also true that a spin 1 color singlet state cannot be produced in gluon fusion. Therefore, we will not consider this case. However, the decay of $X_1\rightarrow ZZ$ is possible and discussed in \cite{25}. \\
Going to $J=2$ resonance, $X_2$, we see that all 5 amplitudes contribute to $gg\rightarrow X_2\rightarrow \gamma \gamma$ process.  In this case however, there are a number of observables with one or both photon's polarization specified, need not be zero. Applying the conditions given in Eq. (14) leads to certain relationships between observables and amplitudes. They are listed in Table 1 for even and odd parity $X_2$ states. In Ref. \cite{11} a massive Kaluza-Klein graviton type coupling is assumed for $X_2 \gamma\gamma$ and $X_2 gg$ vertices and the relevant amplitudes are calculated. They found that $A_1$ amplitude is proportional to $(1+cos\theta)^2$ and $A_4\propto(1-cos\theta)^2$ . This conclusion, except for $\theta=90^\circ$ scattering angle, is inconsistent with our general test given by Eq. (13); namely, for a $J^P=2^+$ resonance one expects to have $A_1=A_4$ at all scattering angles, whereas the results of Ref.\cite{11} shows otherwise. Debate on the possibility of a massive spin 2 resonance is not limited to Ref. \cite{11}; authors of \cite{26} claim that such a state is inconsistent with the existing data and \cite{27} offers yet another method of distinguishing it from the standard model spin-zero Higgs. In Ref. \cite{28} attempts are made to provide a theoretical framework, coming closer to a dynamic independent assessment, but not completely free of dynamical assumptions.\\
Having stated our dynamics independent test, we now can list all the relevant observables and their expressions in terms of the remaining three helicity amplitudes for each resonance state. Choosing the differential cross-section ($\sigma$ in Table 1) to be 1 fixes the overall normalization of all amplitudes. The observables are given in Table 1. \\
It is interesting to note that a subset of this limited set of observables uniquely determine the magnitude of the pertinent amplitudes in each resonance state. The remaining observables (in case of spin-2 resonance) can be used to determine the relative phase of some the amplitudes. \\
Evidently, deviation of any of these observables, excluding $\sigma$, from a null value is an indication that the spin of the observed resonance in $gg\rightarrow \gamma \gamma$ is different from zero, and their sign determine the parity of the resonance state. \\
The test described above can be also used in $q\overline{q}$ collision. However, one should notice that, neglecting orbital angular momentum, in this case the boson states can only have $J=0$ and $J=1$ values.\\

\begin{table}
 {\footnotesize
\begin{tabular}{|c|c|c|c|r|}  \hline
Observable  & $J^P=0^{\pm}$  & $J^P=2^+$  &   $J^P=2^-$ \\ \hline
$\sigma $     & $4 |A_1|^2$    & $4|A_1|^2+8|A_2|^2+4|A_3|^2$   &  $4|A_1|^2+8|A_2|^2+4|A_3|^2$   \\
$(A,A;A,-I)$  & 0   & $4Im(A_1A^{*}_2)$    &$-4Im(A_2 A^{*}_3)$  \\
$(A,A;{\cal{R}}, R)$   &   0 &  $-2 Re (A_1A^{*}_2)$  &  $2 Re (A_2 A^{*}_3)$   \\
$(A,A; R, R)$    &   0   &  $4\{|A_1|^2+|A_3|^2\}$   &  $-4\{|A_1|^2+|A_3|^2\}$   \\
$(A,A;I,-I)$    &   0    &   $ 4\{|A_1|^2+2|A_2|^2-|A_3|^2\}$    &   $ -4\{|A_1|^2-2|A_2|^2-|A_3|^2\}$  \\
$(A,A; \Delta , R)$   &   0   &  $-4 Re(A_1A^{*}_2)$   &    $4 Re(A_2 A^{*}_3)$   \\
\hline
\end{tabular}
\caption{\footnotesize \label{label} Relationship between observables and the bilinear products of helicity amplitudes in the case of resonance with spin-parity $J^P$. First pair of symbols in the argument of each observable refer to the initial gluons, which are identically denoted by A, indicating their unpolarized states. The second pair is pertinent to final state photons polarizations. For further clarification on the notations see the text. }}
\end{table}

\section{Summary and Conclusion}
We have used a set of general criteria to test the direct channel resonance formation.  The so-called $J$-constraint and the factorization constraint establish definite relations between the amplitudes for even and the odd parity states. The established relations between the amplitudes in turn simplifies the observable-amplitude structure of the reaction. We further identified a small set of observables in which only the final state particles' polarization states are specified. A subset of these observables provide insight into the spin and the parity of the resonance. We also note that the assumption of massive Kaluza-Klein type couplings for $x_2$ resonance is inconsistent with our general test. \\

\end{document}